\documentclass[11pt,a4paper,DIV=11,numbers=noenddot]{scrartcl}
\usepackage[utf8]{inputenc}
\makeatletter
\DeclareOldFontCommand{\rm}{\normalfont\rmfamily}{\mathrm}
\DeclareOldFontCommand{\sf}{\normalfont\sffamily}{\mathsf}
\DeclareOldFontCommand{\tt}{\normalfont\ttfamily}{\mathtt}
\DeclareOldFontCommand{\bf}{\normalfont\bfseries}{\mathbf}
\DeclareOldFontCommand{\it}{\normalfont\itshape}{\mathit}
\DeclareOldFontCommand{\sl}{\normalfont\slshape}{\@nomath\sl}
\DeclareOldFontCommand{\sc}{\normalfont\scshape}{\@nomath\sc}
\makeatother
\usepackage{amsmath,amssymb,graphicx,scalefnt,enumitem}
\usepackage[absolute]{textpos}
\usepackage{cite}
\usepackage[affil-it]{authblk}
\usepackage{xcolor}
\usepackage{tabularx,booktabs}
\usepackage{xspace}
\usepackage{listings}
\usepackage[final]{showkeys}
\usepackage{filemod}\usepackage[font=small,labelfont=bf,format=plain,margin=0.05\textwidth]{caption}
\usepackage{subfig}
\usepackage[pdftitle={The light CP-even MSSM Higgs mass including
    N$^3$LO+N$^3$LL QCD corrections},
  pdfauthor={Robert V. Harlander, Jonas Klappert, Alexander Voigt},
  pdfkeywords={Himalaya,H3m,FlexibleSUSY,MSSM,Higgs,supersymmetry},
  bookmarks=true, linktocpage,
  colorlinks=true, allbordercolors=white, allcolors=blue]{hyperref}
\newcommand{\headerline}{{\footnotesize
    TTK--19--40\\
    P3H--19--035\\
    October 2019}}
\newcommand{\fo}{\text{\abbrev{FO}}}

\newcommand{\himalaya}{\code{Himalaya}}

\newcommand{\one}{one}
\newcommand{\two}{two}
\newcommand{\three}{three}
\newcommand{\four}{four}

\newcommand{\lss}{l_{SS}}

\newcommand{\lst}{l_{St}}

\newcommand{\eft}{\abbrev{EFT}}
\newcommand{\ms}{\ensuremath{M_S}}
\newcommand{\msequal}{\ensuremath{\ms^{\text{equal}}}}

\newcounter{notecount}

\newcommand{\citere}[1]{Ref.~\cite{#1}}
\newcommand{\citeres}[1]{Refs.~\cite{#1}}
\newcommand{\code}[1]{\texttt{#1}}
\newcommand{\abbrev}[1]{{\scalefont{.9}#1}}

\newcommand{\eqn}[1]{Eq.\,(\ref{#1})}

\newcommand{\fig}[1]{Fig.\,\ref{#1}}
\newcommand{\figs}[1]{Figs.\,\ref{#1}}
\newcommand{\tab}[1]{Tab.\,\ref{#1}}

\newcommand{\sct}[1]{Sect.\,\ref{#1}}

\newcommand{\dd}{{\rm d}}

\newcommand{\order}[1]{{\cal O}(#1)}

\newcommand{\qcd}{\abbrev{QCD}}
\newcommand{\sm}{\ensuremath{\text{\abbrev{SM}}}}
\newcommand{\mssm}{\ensuremath{\text{\abbrev{MSSM}}}}
\newcommand{\susy}{\abbrev{SUSY}}

\newcommand{\nll}{\abbrev{NLL}}
\newcommand{\nnll}{\abbrev{NNLL}}
\newcommand{\nklo}[1]{\abbrev{N$^{#1}$LO}}
\newcommand{\nkll}[1]{\abbrev{N$^{#1}$LL}}

\newcommand{\msbar}{\ensuremath{\overline{\text{\abbrev{MS}}}}}

\newcommand{\drbarprime}{\ensuremath{\overline{\text{\abbrev{DR}}}^\prime}}

\renewcommand{\Re}{{\rm Re}}

\newcommand{\rge}{\abbrev{RGE}}
\newcommand{\mut}{Q_t}
\newcommand{\mus}{Q_S}
\newcommand{\Qpole}{\mut}
\newcommand{\Qmatch}{\mus}

\newcommand{\msq}[1]{m_{\tilde{q}_{#1}}}
\newcommand{\msu}[1]{m_{\tilde{u}_{#1}}}
\newcommand{\mgluino}{m_{\tilde{g}}}
\newcommand{\mtop}{m_t}
\newcommand{\Mtop}{M_t}
\newcommand{\CP}{\abbrev{CP}}
\newcommand{\unit}[1]{\,\text{#1}}
\newcommand{\MeV}{\unit{MeV}}
\newcommand{\GeV}{\unit{GeV}}
\newcommand{\TeV}{\unit{TeV}}

\newcommand{\FlexibleEFTHiggs}{\textit{Flex\-ib\-le\-EFT\-Higgs}}
\newcommand{\FEFT}{\text{\abbrev{FEFT}}}
\newcommand{\mixed}{\abbrev{\text{hybrid}}}
\newcommand{\HSSUSY}{\code{HSSUSY}}

\newcommand{\FlexibleSUSY}{\abbrev{\texttt{Flex\-ib\-le\-SUSY}}}
\newcommand{\FeynHiggs}{\abbrev{\texttt{Feyn\-Higgs}}}
\newcommand{\SARAH}{\code{SARAH}}
\newcommand{\SPheno}{\code{SPheno}}

\newcommand{\FS}{\code{FS}}

\newcommand{\sinb}[1]{s_\beta^{#1}}
\newcommand{\sintb}[1]{s_{2\beta}^{#1}}
\newcommand{\cosb}[1]{c_\beta^{#1}}

\newcommand{\Mhmixed}{\ensuremath{M_h^\text{hyb}}} 
\newcommand{\MhFO}{\ensuremath{M_h^{\text{\fo}}}} 
\newcommand{\MhFOFS}{\ensuremath{M_h^{\text{\FS}}}} 
\newcommand{\MhEFT}{\ensuremath{M_h^{\text{\eft}}}} 
\newcommand{\MhFEFT}{\ensuremath{M_h^{\FEFT}}}

\newcommand{\Deltasupp}{\ensuremath{\Delta_{v}}}
\newcommand{\Deltabarsupp}{\ensuremath{\bar{\Delta}_{v}}}
\newcommand{\slha}{\abbrev{SLHA}}
\newcommand{\Mathematica}{\code{Mathematica}}
\newcommand{\DeltaTL}{\ensuremath{\Delta^{2\ell}_{\order{y_t^4 g_3^2 + y_t^6}}}}
\newcommand{\DeltaTLAtAs}{\ensuremath{\Delta^{2\ell}_{\order{y_t^4 g_3^2}}}}
\newcommand{\DeltaTLAtAt}{\ensuremath{\Delta^{2\ell}_{\order{y_t^6}}}}
\newcommand{\DeltaTLSupp}{\ensuremath{\left.\DeltaTL\right|_{v^2\ll\ms^2}}}
\newcommand{\DeltaTLSuppAtAs}{\ensuremath{\left.\DeltaTLAtAs\right|_{v^2\ll\ms^2}}}
\newcommand{\DeltaTLSuppAtAt}{\ensuremath{\left.\DeltaTLAtAt\right|_{v^2\ll\ms^2}}}
\newcommand{\DeltaYtG}{\ensuremath{\Delta^{1\ell}_{(g_3^2)} y_t}}
\newcommand{\DeltaYtYt}{\ensuremath{\Delta^{1\ell}_{(y_t^2)} y_t}}
\newcommand{\DeltaLaYtG}{\ensuremath{\Delta^{2\ell}_{(y_t^4 g_3^2)} \lambda}}
\newcommand{\DeltaLaYtYt}{\ensuremath{\Delta^{1\ell}_{(y_t^4)} \lambda}}
\newcommand{\DeltaLaYtYtYt}{\ensuremath{\Delta^{2\ell}_{(y_t^6)} \lambda}}
\newcommand{\DeltaVYt}{\ensuremath{\Delta^{1\ell}_{(y_t^2)} v}}

\title{The light CP-even MSSM Higgs mass including
    N$^\mathbf{3}$LO+N$^\mathbf{3}$LL QCD corrections}
\author{R.V. Harlander, J. Klappert, and A. Voigt}
\affil{Institute for Theoretical Particle Physics and Cosmology,\\ RWTH
  Aachen University, 52074 Aachen, Germany}
\date{}

\begin{document}
\maketitle
\thispagestyle{empty}
\begin{abstract}
  We present a calculation of the light neutral \CP-even Higgs boson
  pole mass in the real \mssm\ which combines state-of-the-art \eft\ and
  fixed-order results, including the \three-loop fixed-order
  \qcd\ corrections as well as the resummation of logarithmic terms in
  the ratio of the weak to the \susy\ scale up to fourth logarithmic
  order. This hybrid calculation should be valid for arbitrary
  \susy\ scales above the weak scale. Comparison to the pure fixed-order and
  \eft\ results provides an estimate of their individual validity range.
\end{abstract}

\begin{textblock*}{10em}(\textwidth,1.5cm)
\raggedright\noindent
\headerline
\end{textblock*}

\clearpage
\tableofcontents
\clearpage

\section{Introduction}

With the discovery of the Higgs boson with a mass of $ M_h = (125.10\pm
0.14) \GeV $
\cite{Aad:2012tfa,Chatrchyan:2012xdj,Aad:2015zhl,Tanabashi:2018oca}, the
Standard Model (\sm) of particle physics is complete and appears to be a
good description of nature around and below the electroweak scale.
However, the \sm\ does not describe gravity and cannot account for
phenomena typically associated with dark matter, for example, or for
\CP-violation at the level required to explain the observed baryon
anti-baryon asymmetry. Supersymmetry (\susy) has been an attractive
proposal to address some of the deficits of the \sm.  One particular
feature of the Minimal Supersymmetric Standard Model (\mssm) is its
constrained Higgs sector which, for a given set of \susy\ parameters,
results in a theoretical value of the lightest Higgs boson
mass. Comparison to the measured mass of the observed Higgs boson as
quoted above provides a stringent constraint of the \mssm.  It is well
known that this theory value of the lightest Higgs boson mass receives
large radiative corrections, so that higher-order calculations are
required in order to achieve a precision which is competitive with the
experimental accuracy.

There are different methods to calculate the Higgs pole mass in the
\mssm, which can be divided into fixed-order (\fo), effective field
theory (\eft) and hybrid approaches.
In the fixed-order calculation, loop corrections to the Higgs mass are
calculated in the full \mssm, and the perturbation series is truncated
at a fixed order of the coupling constants.  If the \susy\ particles
have masses not too far above the electroweak scale, the
\fo\ calculation typically leads to a reliable value.  However, if (some
of) the \susy\ particles are very heavy, then the perturbative
coefficients receive large logarithmic contributions, which spoil the
perturbative series. Currently, loop corrections up to the \two-loop
level are known in the on-shell scheme
\cite{Hempfling:1993qq,Heinemeyer:1998kz,Heinemeyer:1998jw,Heinemeyer:1998np,
  Heinemeyer:1999be,Degrassi:2001yf,Brignole:2001jy,Brignole:2002bz,
  Dedes:2003km,
  Heinemeyer:2004xw,Heinemeyer:2007aq,Degrassi:2014pfa,Borowka:2014wla,
  Hollik:2014bua,
  Borowka:2015ura,Passehr:2017ufr,Borowka:2018anu}
and up to \three-loop level in the \drbarprime\ scheme
\cite{Degrassi:2001yf,Brignole:2001jy,Martin:2001vx,Martin:2002iu,
  Martin:2002wn,Dedes:2002dy,Brignole:2002bz,Dedes:2003km,Martin:2003it,
  Allanach:2004rh,Martin:2004kr,Martin:2005eg,Martin:2007pg,Harlander:2008ju,
  Kant:2010tf,Degrassi:2014pfa,Goodsell:2016udb,Martin:2017lqn,
  Harlander:2017kuc,R.:2019ply,
  Goodsell:2019zfs}.
The corresponding \fo\ Higgs pole mass results are available through
implementations into publicly available spectrum generators
\cite{Heinemeyer:1998yj,Heinemeyer:1998np,Degrassi:2002fi,Frank:2006yh,
  Hahn:2013ria,Allanach:2001kg,Djouadi:2002ze,Porod:2003um,Porod:2011nf,
  Athron:2014yba,Athron:2017fvs}.

An \eft\ calculation, on the other hand, is based on the assumption that
the \susy\ particles are very heavy compared to the electro-weak scale.
Integrating them out leaves the \sm\ as an \eft. The latter retains the
\susy\ constraints through the matching conditions between the
\mssm\ and the \sm\ parameters, which are imposed at some large mass
scale.  The Higgs pole mass is then calculated from the
\sm\ \msbar\ parameters after evolving them down to the electro-weak
scale through \sm\ renormalization group equations (\rge{}s), thereby
resumming contributions which are logarithmic in the ratio of the
\susy\ and the electro-weak scale (``large logarithms''). This procedure
has been implemented through third logarithmic order (\nnll) in several
publicly available pure-\eft\ spectrum generators
\cite{Vega:2015fna,Athron:2017fvs,Lee:2015uza}.  Resummation through
fourth logarithmic order (\nkll{3}) has recently been achieved through the
evaluation of the \three-loop matching coefficient for the quartic
coupling \cite{Harlander:2018yhj}, which complemented the available
\two-loop matching relations
\cite{Draper:2013oza,Bagnaschi:2014rsa,Lee:2015uza,Vega:2015fna,
  Bagnaschi:2017xid,Bagnaschi:2019esc,Murphy:2019qpm
}.

It turns out that, in order for the theoretical value of the light
\mssm\ Higgs mass to be compatible with the observed Higgs mass of
$M_h\approx 125$\,GeV, the \susy\ spectrum requires TeV-scale stops (see
\citeres{Bagnaschi:2014rsa,Vega:2015fna,Bahl:2016brp,
  Bahl:2017aev,Allanach:2018fif}, for example).  It is not clear
\textit{a priori} whether a \fo\ or an \eft\ approach provides the best
value for the Higgs mass at these mass scales.  For this reason,
so-called hybrid approaches have been devised
\cite{Hahn:2013ria,Bahl:2016brp,Athron:2016fuq,Bahl:2017aev,Athron:2017fvs,Staub:2017jnp,Bahl:2018jom}.
They combine the virtues of a \fo\ and an \eft\ calculation, and lead to
a reliable value for the Higgs pole mass at all \susy\ scales.  So far,
they rely on \two-loop \fo\ results with a resummation of the large
logarithms at \nnll\ level at most. Comparison to the highest available
\fo\ result shows good agreement up to remarkably large \susy\ scales of
the order of $5$--$10\TeV$ \cite{R.:2019irs}, in accordance with earlier
comparisons of \fo\ and \eft\ results \cite{Harlander:2017kuc}.

In this paper, we adopt a hybrid approach for the real \mssm\ by
including the next perturbative order in the strong coupling. More
precisely, we combine the \fo\ and the \eft\ results at order $y_t^4
g_3^4$, resulting in a value at fourth perturbative order (\nklo{3})
with \nkll{3}\ resummation.\footnote{Note that, in our notation, the
  one-loop top-quark induced corrections to the light \mssm\ Higgs mass
  are of order $y_t^4$, while other authors would denote them as
  $\order{\alpha_t}$ (see, e.g.,
  \citeres{Degrassi:2001yf,Brignole:2001jy}).} By comparing our
\three-loop hybrid result with the individual \fo\ and
\eft\ approximations, we infer the size of the terms of
$\order{v^2/\ms^2}$ which are usually neglected in a pure
\eft\ approach. This allows us to derive an estimate for the
\susy\ scale above which a pure \eft\ calculation is sufficient (see
also \citere{Allanach:2018fif}).

The remaining part of this paper is structured as follows: In
\sct{sec:matching} we describe our procedure to combine the \three-loop
\fo\ and \eft\ results. The numerical implications of the resulting
hybrid result are discussed in \sct{sec:results}. \sct{sec:conclusions}
contains our conclusions.

\section{Matching procedure}
\label{sec:matching}

\subsection{General outline}

So far, two approaches to combine \fo\ and \eft\ results in the context
of the \susy\ Higgs mass have been pursued in the literature:
\begin{itemize}
\item \textbf{Subtraction approach}: Here one writes the squared Higgs
  pole mass as
  \begin{align}
    (M_h^{\text{subtr}})^2 = (M_h^{\fo})^2 - (M_h^{\text{logs}})^2 + (M_h^{\text{res}})^2,
    \label{eq:FH_matching}
  \end{align}
  where $(M_h^{\fo})^2$ denotes the \fo\ result,
  $(M_h^{\text{logs}})^2$ are the large logarithmic \fo\
  corrections, and $(M_h^{\text{res}})^2$ are the resummed logarithmic
  corrections.

  An advantage of this approach is that existing fixed-order results can
  be used and different effective theories can be considered in a
  straightforward way.  The generalization of this approach to models
  beyond the \mssm\ is non-trivial, because it requires model-specific
  \fo\ and \eft\ loop calculations.

  This approach is implemented in \FeynHiggs\ at the \two-loop level,
  for example \cite{Hahn:2013ria,Bahl:2016brp,Bahl:2017aev}.

\item \textbf{FlexibleEFTHiggs approach}
  \cite{Athron:2016fuq,Athron:2017fvs,Staub:2017jnp}: Here one employs
  the identity
  \begin{align}
    (M_h^\sm)^2 = (M_h^\mssm)^2\,,
    \label{eq:FEFT_matching}
  \end{align}
  where $M_h^\sm$ denotes the Higgs pole mass expressed through
  \sm\ \msbar\ parameters, and $M_h^\mssm$ is the Higgs pole mass
  calculated in the \mssm\ in the \drbarprime\ scheme. The \msbar\ and
  \drbarprime\ parameters appearing in \eqn{eq:FEFT_matching} depend on
  the renormalization scale $\mus$, which is set close to the
  \susy\ scale.  This determines the \sm\ quartic Higgs coupling in the
  \msbar\ scheme at the scale $\mus$, which is then evolved down to the
  electro-weak scale using \sm\ \rge{}s in order to evaluate the Higgs
  pole mass from it.

  Due to the simplicity of the matching condition
  \eqref{eq:FEFT_matching}, this approach can be generalized to other
  models in a rather straightforward way.  However, the extension of the
  approach to the \two-loop level is non-trivial, because care must be
  taken to cancel potential large logarithmic corrections in the
  matching.

  The \FlexibleEFTHiggs\ approach is implemented at \one-loop level into
  \FlexibleSUSY\ \cite{Athron:2016fuq,Athron:2017fvs}, and at \two-loop
  level into \SARAH/\SPheno\ \cite{Staub:2017jnp}.\footnote{Note that in
    the implementation of the \FlexibleEFTHiggs\ approach in
    \SARAH/\SPheno, large higher-order logarithmic corrections are
    induced at the matching scale.  As a result, \SARAH/\SPheno\ resums
    large logarithms only up to (including) the leading-log level.}
\end{itemize}

In this paper, we adopt a hybrid scheme which is similar to the
subtraction approach of \eqn{eq:FH_matching}.  However, we work in the
\drbarprime\ scheme and go one loop level higher, combining the \fo\ and
\eft\ approximations which include \three-loop level \qcd\ corrections.
In our scheme, we calculate the (squared) Higgs pole mass as
\begin{align}
  (\Mhmixed)^2 = (\MhEFT)^2 + \Deltasupp,
  \label{eq:Mh_mixed}
\end{align}
where $\MhEFT$ denotes the \three-loop \eft\ result of
\FlexibleSUSY/\HSSUSY+\himalaya\ \cite{Harlander:2018yhj}; it resums
large logarithms of order $y_t^4g_3^6$ to \nkll{3}, while others are
resummed to \nnll.\footnote{Our identification of the logarithmic order
  refers to the required order of the $\beta$ function of the \sm\ Higgs
  self coupling $\lambda$. Specifically, our \nkll{n} terms involve the
  $\beta$ function to $\order{y_t^4g_3^{2n}}$.}  Its fixed-order
expansion would reproduce the full fixed-order result in the limit
$v^2/\ms^2\to 0$, including the known \two-loop corrections in the
gaugeless limit and the \three-loop terms of order $y_t^4g_3^4$ from
\himalaya \cite{Kant:2010tf,Harlander:2008ju,Harlander:2017kuc},
including non-logarithmic terms.  $\Deltasupp$ supplies the terms that
are suppressed by powers of $v^2/\ms^2$ as $\ms\gg v$ at fixed order up
to the \two-loop level. We separate \Deltasupp\ into a
tree-level+\one-loop and a \two-loop part,
\begin{align}
  \Deltasupp &= \Deltasupp^{0\ell+1\ell} + \Deltasupp^{2\ell}\,.
\end{align}
These terms are extracted from the \FlexibleEFTHiggs\ result implemented
in \FlexibleSUSY, and from the \two-loop contributions included in the
\himalaya\ library, as described in what follows.
The tree-level and \one-loop contribution $\Deltasupp^{0\ell+1\ell}$ is
obtained by taking the difference between the \one-loop
\FlexibleEFTHiggs\ result $\MhFEFT$ and the \one-loop pure \eft\ result
obtained from \HSSUSY\ as
\begin{align}
  \Deltasupp^{0\ell+1\ell} &= \left[(\MhFEFT)^2 - (\MhEFT)^2\right]_{0\ell+1\ell}.
  \label{eq:Delta_supp_1L}
\end{align}
Due to the structure of the \FlexibleEFTHiggs\ calculation, this
difference contains all tree-level and \one-loop \susy\ contributions of
higher order in $v^2/\ms^2$, and formally \two-loop non-logarithmic
electroweak \susy\ terms (see below).  In particular, large logarithmic
corrections as well as \two-loop non-electroweak \susy\ contributions
are absent.
The \two-loop contribution $\Deltasupp^{2\ell}$ is obtained as
\begin{align}
  \Deltasupp^{2\ell} &= \DeltaTL - \DeltaTLSupp.
  \label{eq:Delta_supp_2L}
\end{align}
The terms on the r.h.s.\ of \eqn{eq:Delta_supp_2L} represent the
difference between the \two-loop fixed-order contribution $\order{y_t^4
  g_3^2 + y_t^6}$ calculated with \himalaya, and the same \two-loop
fixed-order contribution where all $\order{v^2/\ms^2}$ terms are
neglected.  This difference thus contains all \two-loop
$\order{v^2/\ms^2}$ terms at $\order{y_t^4 g_3^2 + y_t^6}$.  Large
logarithmic as well as non-electroweak \three-loop corrections of order
$(v^2/\ms^2)^0$ are absent.

\subsection{Explicit result in the degenerate-mass case}

\himalaya\ includes $\DeltaTLSupp$ for a general \susy\ spectrum, but we
find that the full analytic expression is too long to be displayed in
this paper. For reference, however, we include the results for the case
of degenerate \drbarprime\ \susy\ mass parameters at the \susy\ scale
$\ms$, i.e., $\msq{3}(\ms) = \msu{3}(\ms) = \mgluino(\ms) = m_A(\ms) =
\mu(\ms) = \ms$.  Here $\msq{3}$ and $\msu{3}$ denote the left- and
right-handed third generation squark mass parameters, $\mgluino$ the
gluino mass, $m_A$ the \CP-odd Higgs boson mass, and $\mu$ the
superpotential $\mu$-parameter.  We parameterize our expressions in
terms of the \drbarprime\ stop mixing parameter $X_t = A_t -
\mu/\tan\beta$, where $A_t$ is the trilinear Higgs-stop-stop coupling,
and $\tan\beta = v_u/v_d$ with $v_u$ and $v_d$ being the running vacuum
expectation values of the \mssm\ up- and down-type Higgs doublets,
respectively.  Furthermore we define the short-hand notation $x_t =
X_t/\ms$, $\sinb{} = \sin\beta$, $\cosb{} = \cos\beta$, $\sintb{} = \sin
2\beta$, $\kappa = 1/(4\pi)^2$, $\lss = \log(M_S^2/\mus^2)$ and $\lst =
\log(M_S^2/\mtop^2)$.  Here, $\mus$ denotes the renormalization scale at
which the matching is performed, $\mtop = y_t \sinb{} v / \sqrt{2}$ is
the running top quark mass, $y_t$ denotes the top Yukawa coupling, $g_3$
is the strong gauge coupling, and $v = (v_u^2 + v_d^2)^{1/2}$ is the
SM-like Higgs vacuum expectation value, all defined in the \mssm\ in the
\drbarprime\ scheme.

Following the procedure described in \citere{Harlander:2018yhj}, the
\two-loop subtraction term on the r.h.s.\ of \eqn{eq:Delta_supp_2L}
can be expressed in terms of threshold corrections as
\begin{align}
  \DeltaTLSupp &= \DeltaTLSuppAtAs + \DeltaTLSuppAtAt
\end{align}
with
\begin{align}
  \DeltaTLSuppAtAs &=
  \frac{\kappa^2 y_t^4g_3^2v^2\sinb{4}}{2}\Bigg\{
  \DeltaLaYtG - 8\Big[\DeltaYtG\left(3+6\lss - 6\lst\right)\nonumber\\
  &~~~~~~~~~~~~~~~~~~~~~- 4\left(1+3\lss-3\lst\right)\left(\lss-\lst\right)\Big]
    \Bigg\}\, ,
  \\
  \DeltaTLSuppAtAt &=
    \frac{\kappa^2 y_t^6 v^2 \sinb{4}}{2}\Bigg\{
    4 \DeltaYtYt \left(-6 + \DeltaLaYtYt + 12 \lst - 12 \lss\right) \nonumber \\
    &~~~~~~~~~~~~~~~~~~ + \sinb{2} \Big[-12 + 2 \DeltaLaYtYtYt
    + \DeltaLaYtYt \left(2 + 2 \DeltaVYt - 3 \lst + 3\lss\right) \nonumber \\
    &~~~~~~~~~~~~~~~~~~~~~~~~~~ + 24 \DeltaVYt (\lst - \lss - 1) \nonumber \\
    &~~~~~~~~~~~~~~~~~~~~~~~~~~- 18 (\lst - \lss) (1 + 3 \lst - 3 \lss)
    - 2 \pi^2\Big]\Bigg\}\,,
\end{align}
and the \one- and \two-loop threshold factors
\cite{Martin:2007pg,Bagnaschi:2014rsa}
\begin{subequations}
\begin{align}
  \DeltaYtG &= \frac{4}{3}\left(1 + \lss  - x_t\right),\\
  \DeltaYtYt & = \frac{1}{8}\left[6\lss + \cosb{2}\left(-3 + 6\lss\right) - 2\sinb{2}x_t^2\right],\\
  \DeltaVYt &= \frac{x_t^2}{4},\\
  \DeltaLaYtYt &= 12 \lss + 12x_t^2 - x_t^4,\\
  \DeltaLaYtG & = -\frac{8}{3}\left[12 \lss^2 + x_t\left(24 - 12x_t - 4x_t^2 + x_t^3\right)
    + 8\lss \left(-3-3 x_t + 3x_t^2+x_t^3\right)\right].
\end{align}
\end{subequations}
The \two-loop correction $\DeltaLaYtYtYt$ is given by Eq.~(21) of
\citere{Vega:2015fna} (without the prefactor).  Inserting the threshold
corrections, the subtraction terms become
\begin{align}
    \DeltaTLSuppAtAs &= \kappa^2y_t^4g_3^2 \sinb{4} v^2\Bigg[
    -16 + 16\lst - 64\lss\lst + 48\lst^2 \nonumber \\
    &~~~~~~~~~~~~~~~~~~~~~~ + x_t\left(-16 + 64\lss - 32\lst\right)
    + x_t^2\left(16 - 32\lss\right) \nonumber \\
    &~~~~~~~~~~~~~~~~~~~~~~ + \frac{x_t^3\left(16 - 32\lss\right)}{3}
    - \frac{4x_t^4}{3}\Bigg], \\
    \DeltaTLSuppAtAt &=
    \frac{\kappa^2y_t^6\sinb{2}v^2}{16}\Bigg\{-144 - 768K + 144\lss \nonumber \\
    &\qquad + \sintb{2}\Big(48 + 720K - 36\lss - 36\lst + 144\lss\lst + 12\pi^2\Big) \nonumber \\
    &\qquad + \sinb{4}\Big(336 - 288\lss - 144\lst + 864\lss\lst - 432\lst^2 - 16\pi^2\Big) \nonumber \\
    &\qquad - x_t\sintb{}\left(432 + 2304K - 432\lss\right) \nonumber \\
    &\qquad - x_t^2\Big[-224 - 1152K + 144\lss + \sintb{2}\left(228 + 1248K - 432\lss\right) \nonumber \\
    &\qquad ~~~~~~~~ +\sinb{4}\left(576 - 2160\lss + 336\lst\right)\Big] \nonumber \\
    &\qquad -x_t^3\sintb{}\left(-320 - 1536K + 192\lss\right) \nonumber \\
    &\qquad -x_t^4\Big[76 + 384K - 24\lss - \sintb{2}\left(110 + 480K - 90\lss\right) \nonumber \\
    &\qquad ~~~~~~~~ -\sinb{4}\left(248 - 408\lss + 24\lst\right)\Big] \nonumber \\
    &\qquad -x_t^5\sintb{}\left(76 + 384K - 24\lss\right) \nonumber \\
    &\qquad -x_t^6\Big[\sintb{2}\left(19 + 96K - 6\lss\right) + \sinb{4}\left(20 - 24\lss\right)\Big]\Bigg\},
\end{align}
where $K = -\sqrt{1/3}\int_0^{\pi/6}\dd x\ln(2\cos x)\approx
-0.1953256$.

\section{Numerical results}
\label{sec:results}

\subsection[Size of the $\order{v^2/\ms^2}$ terms]{Size of the
  \boldmath{$\order{v^2/\ms^2}$} terms}
\label{sec:supp}

\begin{figure}[tb!]
  \centering
  \includegraphics[width=0.49\textwidth]{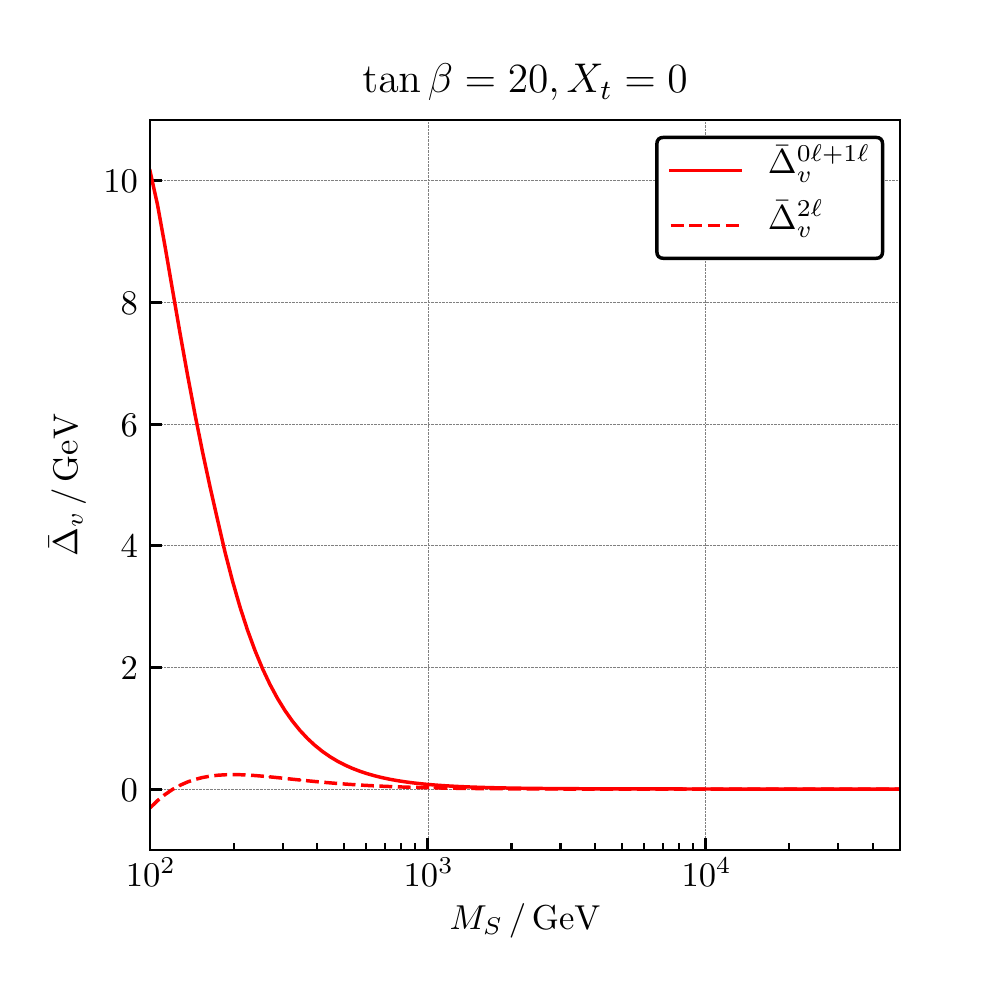}\hfill
  \includegraphics[width=0.49\textwidth]{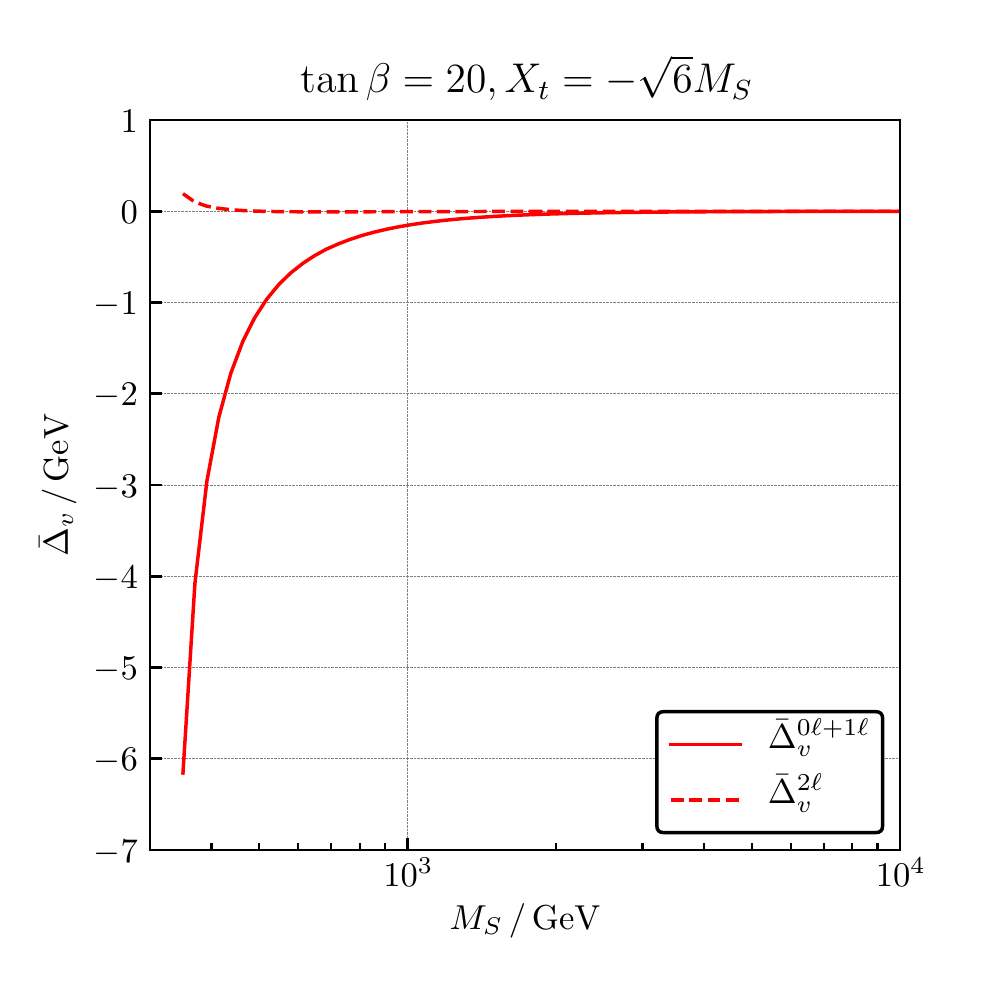}
  \includegraphics[width=0.49\textwidth]{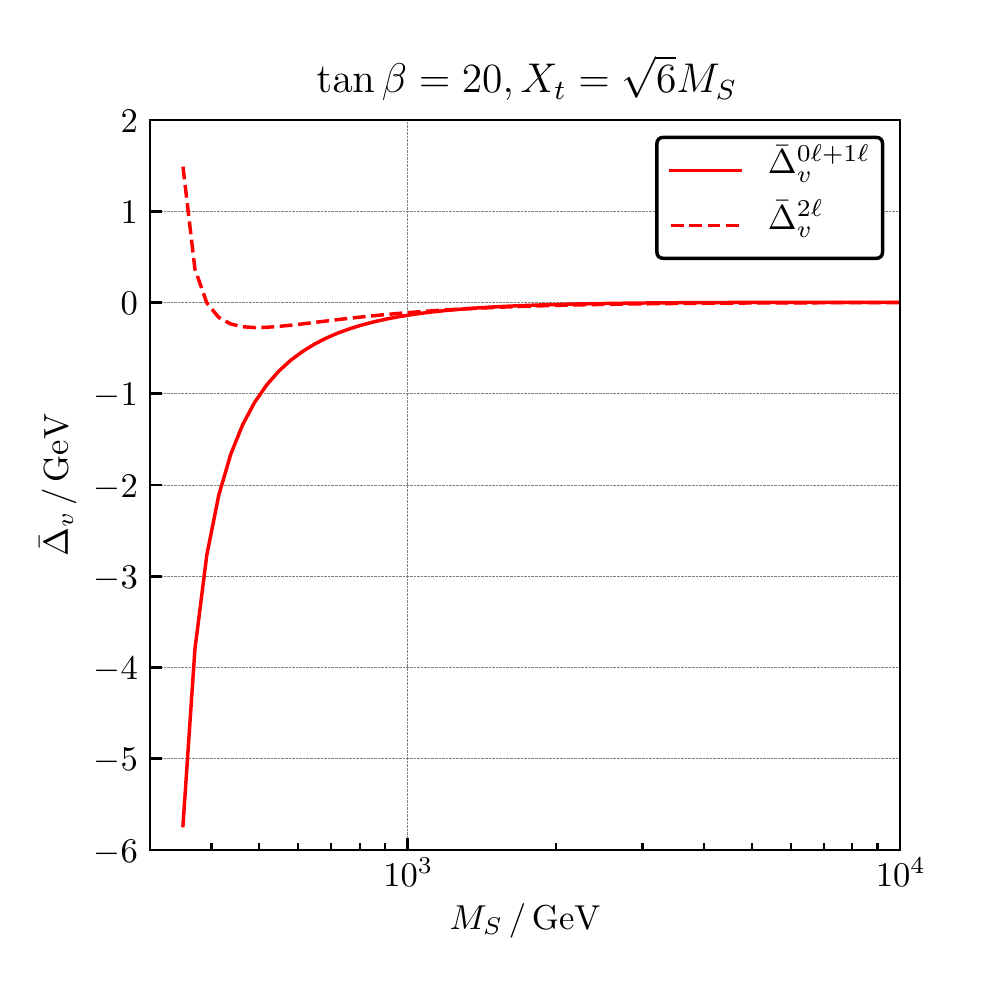}
  \caption{Size of the $\order{v^2/\ms^2}$ terms.}
  \label{fig:scan_MS_supp}%
\end{figure}

In this section, we study the effect of the $\order{v^2/\ms^2}$ terms
$\Deltasupp$ on the Higgs pole mass as a function of the
\susy\ scale. Even though our approach is applicable to a general
\susy\ mass spectrum, we focus on the degenerate mass case in our
numerical examples.  For convenience we define the (non-squared)
contribution of these terms as
\begin{subequations}
  \begin{align}
  \Deltabarsupp &= \Deltabarsupp^{0\ell+1\ell} + \Deltabarsupp^{2\ell}, \\
  \Deltabarsupp^{0\ell+1\ell} &=
  \left[(\MhEFT)^2 + \Deltasupp^{0\ell+1\ell}\right]^{1/2} - \MhEFT , \\
  \Deltabarsupp^{2\ell} &=
  \left[(\MhEFT)^2 + \Deltasupp^{0\ell+1\ell}  + \Deltasupp^{2\ell}\right]^{1/2}
  - \left[(\MhEFT)^2 + \Deltasupp^{0\ell+1\ell}\right]^{1/2}\,.
\end{align}
\end{subequations}
Setting $\tan\beta = 20$, we find that the $\order{v^2/\ms^2}$ terms can
be sizable below $\ms \lesssim 0.5\TeV$, 
while they are small as long as $\ms\gtrsim 1$\,TeV, see
\fig{fig:scan_MS_supp}. Specifically, we find for $\tan\beta = 20$ and
$\ms\gtrsim 1$\,TeV:
\begin{subequations}
  \begin{alignat}{2}
    x_t &= 0        &\ :\qquad  |\Deltabarsupp| &\lesssim 0.10\GeV,\\
    x_t &= -\sqrt{6}&\ :\qquad  |\Deltabarsupp| &\lesssim 0.15\GeV,\\
    x_t &=  \sqrt{6}&\ :\qquad  |\Deltabarsupp| &\lesssim 0.25\GeV.
  \end{alignat}
\end{subequations}
Other values of $\tan\beta$ lead to similar observations.

The sign and the order of magnitude of these results are in agreement with
the contribution due to higher-dimensional operators as presented in
\citere{Bagnaschi:2017xid}.  Since the remaining uncertainty on the
Higgs pole mass is dominated by the uncertainty induced by the
extraction of the running top Yukawa coupling, which has been estimated
to be between $0.2$--$0.6\GeV$
\cite{Buttazzo:2013uya,Vega:2015fna,Bagnaschi:2017xid,Allanach:2018fif},
we conclude that for $\ms \gtrsim 1\TeV$ the $\order{v^2/\ms^2}$ terms
are negligible and the \eft\ approach leads to a more precise value of
the Higgs pole mass than the fixed-order result.  These findings are
compatible with the transition region of $\msequal = 1.0$--$1.3\TeV$
estimated in \citere{Allanach:2018fif}.

\subsection{Comparison of fixed-order, \eft, and \mixed\ results}

\subsubsection{Convergence for high \susy\ scales}
\label{sec:convergence}

\begin{figure}[tbh!]
  \centering
  \subfloat[][]{%
    \includegraphics[width=0.49\textwidth]{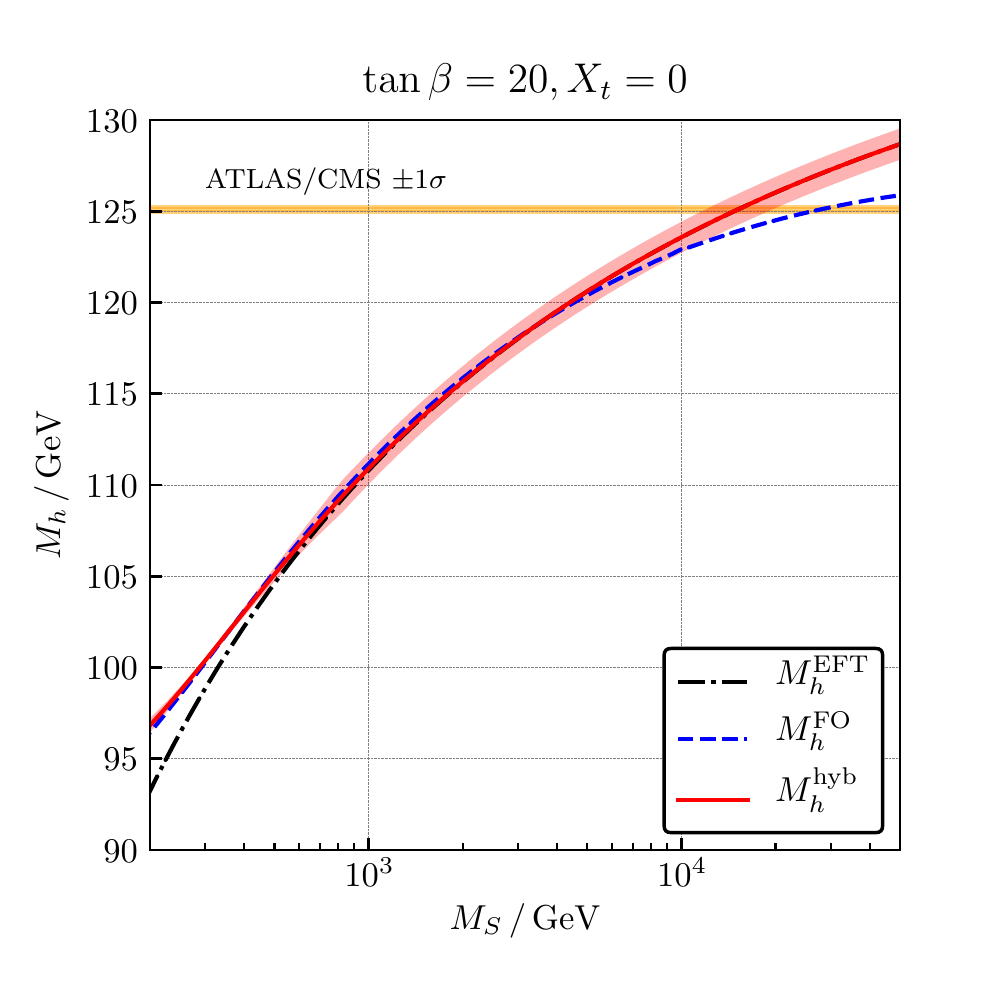}%
    \label{fig:3L_Mh_abs_Xt-0}%
  }\hfill
  \subfloat[][]{%
    \includegraphics[width=0.49\textwidth]{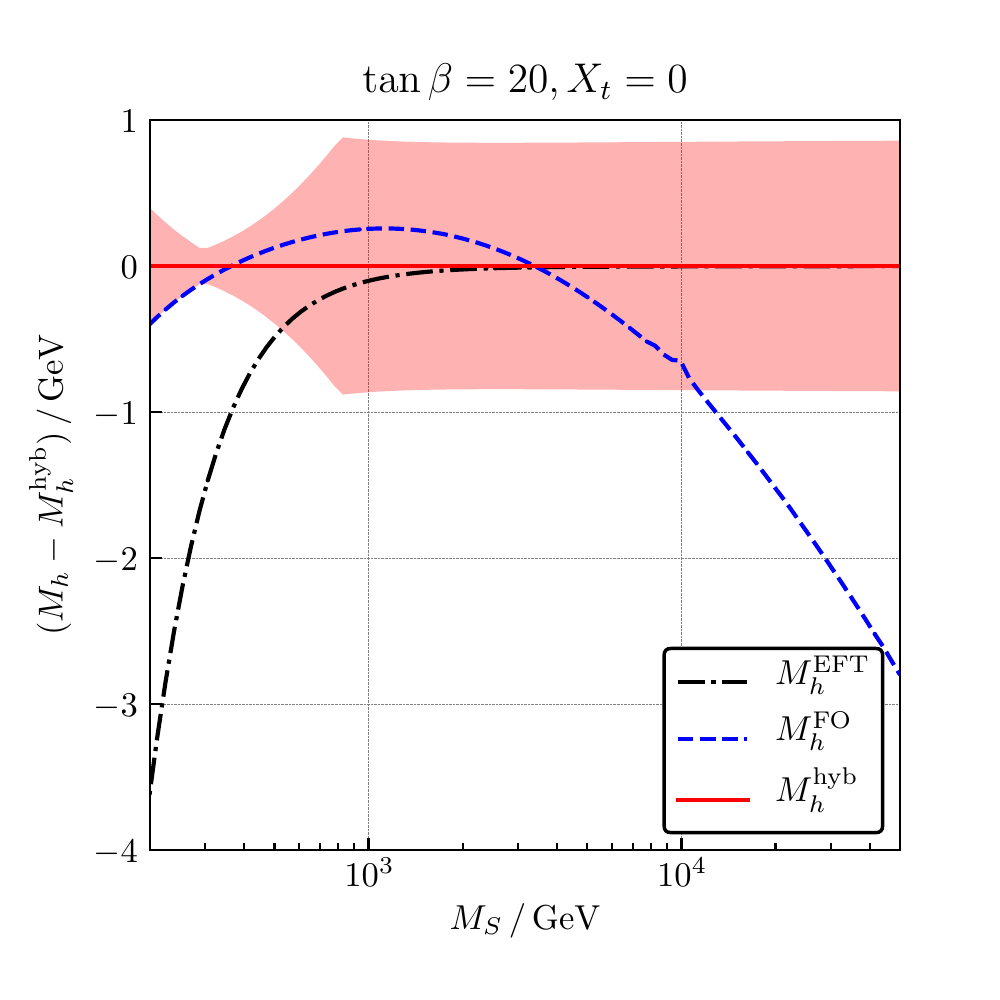}%
    \label{fig:3L_Mh_diff_Xt-0}%
  }\\
  \subfloat[][]{%
    \includegraphics[width=0.49\textwidth]{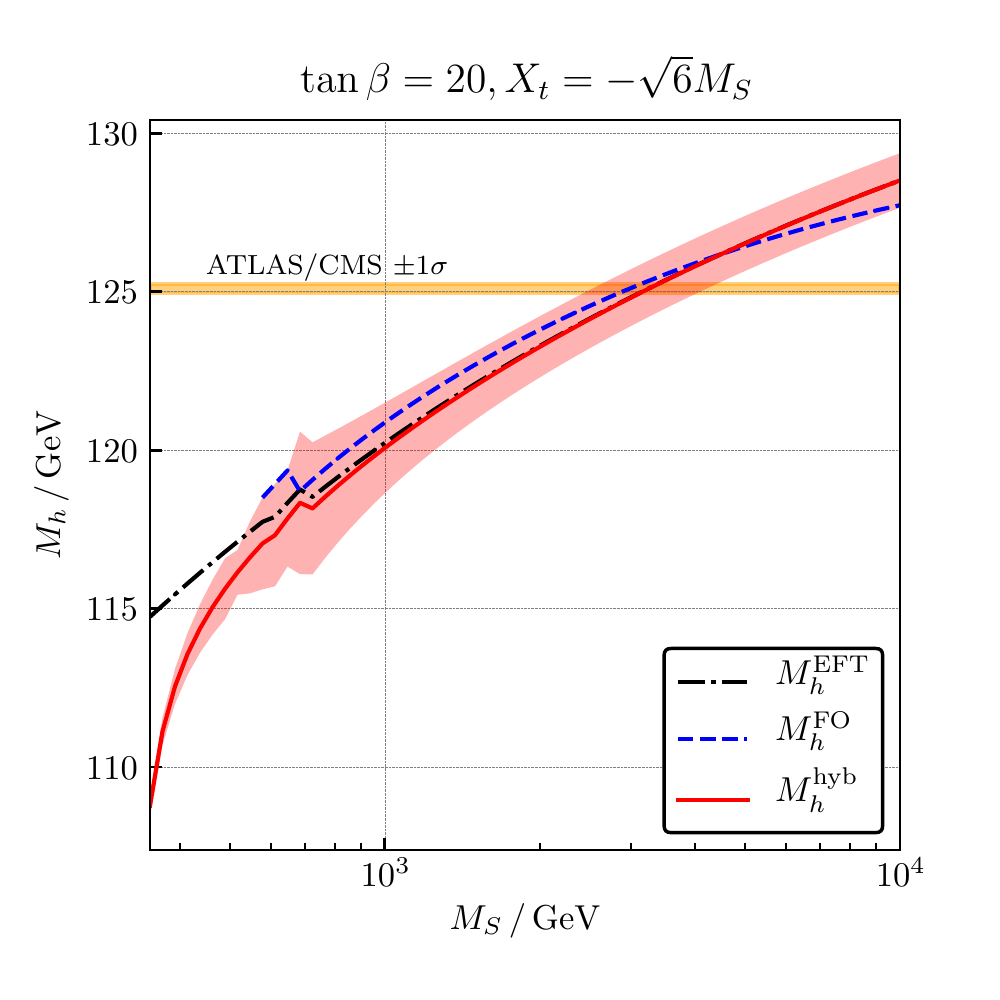}%
    \label{fig:3L_Mh_abs_Xt--sqrt6}%
  }\hfill
  \subfloat[][]{%
    \includegraphics[width=0.49\textwidth]{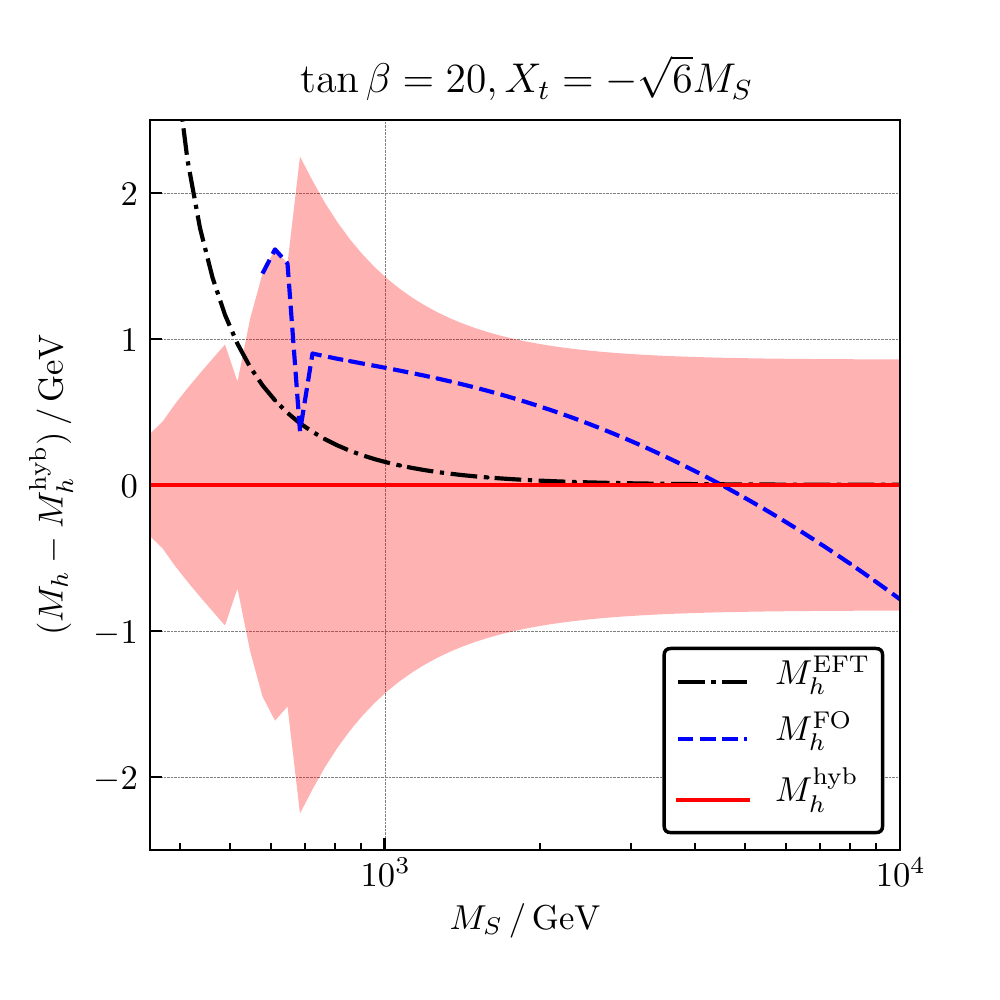}%
    \label{fig:3L_Mh_diff_Xt--sqrt6}%
  }
  \caption{Comparison of the \three-loop \fo,
    \eft, and \mixed\ results.}
  \label{fig:scan_MS_mixed_3L}
\end{figure}

In \fig{fig:scan_MS_mixed_3L}, we compare the \mixed\ result defined
through \eqn{eq:Mh_mixed} (red solid line) with the \three-loop
\drbarprime\ fixed-order approximation $\MhFO$ of
\FlexibleSUSY+\himalaya\ \cite{Harlander:2017kuc} (blue dashed line) and
the \three-loop \eft\ result $\MhEFT$ of
\FlexibleSUSY/\HSSUSY+\himalaya\ \cite{Harlander:2018yhj} (black
dash-dotted line), which resums large logarithms through \nklo{3}. The
red band is our uncertainty estimate on the \mixed\ result (see
\sct{sec:uncertainty} below for details).  Since $\Deltasupp\to 0$ for
$\ms\to\infty$, the \mixed\ curve converges towards the \eft\ curve in
this limit. Note that in the scenario with $x_t = -\sqrt{6}$ for values
of $M_S$ below $\sim600$ GeV, no suitable mass hierarchy is available in
\himalaya.  The \three-loop fixed-order contribution is set to zero in
this case, which means that the \eft\ curve and the \mixed\ calculation
is formally consistent only at the \two-loop level for lower scales.
On the other hand, for $\ms\to M_Z$ one may expect the \mixed\ curve to
converge towards the \three-loop fixed-order curve.  However, we find a
finite offset at low energies of up to $\sim 0.5\GeV$ for $x_t = 0$ and
$\sim 1.5\GeV$ for $x_t = -\sqrt{6}$.  This offset results from higher
order $\order{v^2/\ms^2}$ terms, which are not suppressed in the low
$\ms$ region.  The origin of these will be investigated in the following
sub-section.

\begin{figure}[tbh]
  \centering
  \subfloat[][]{%
    \includegraphics[width=0.49\textwidth]{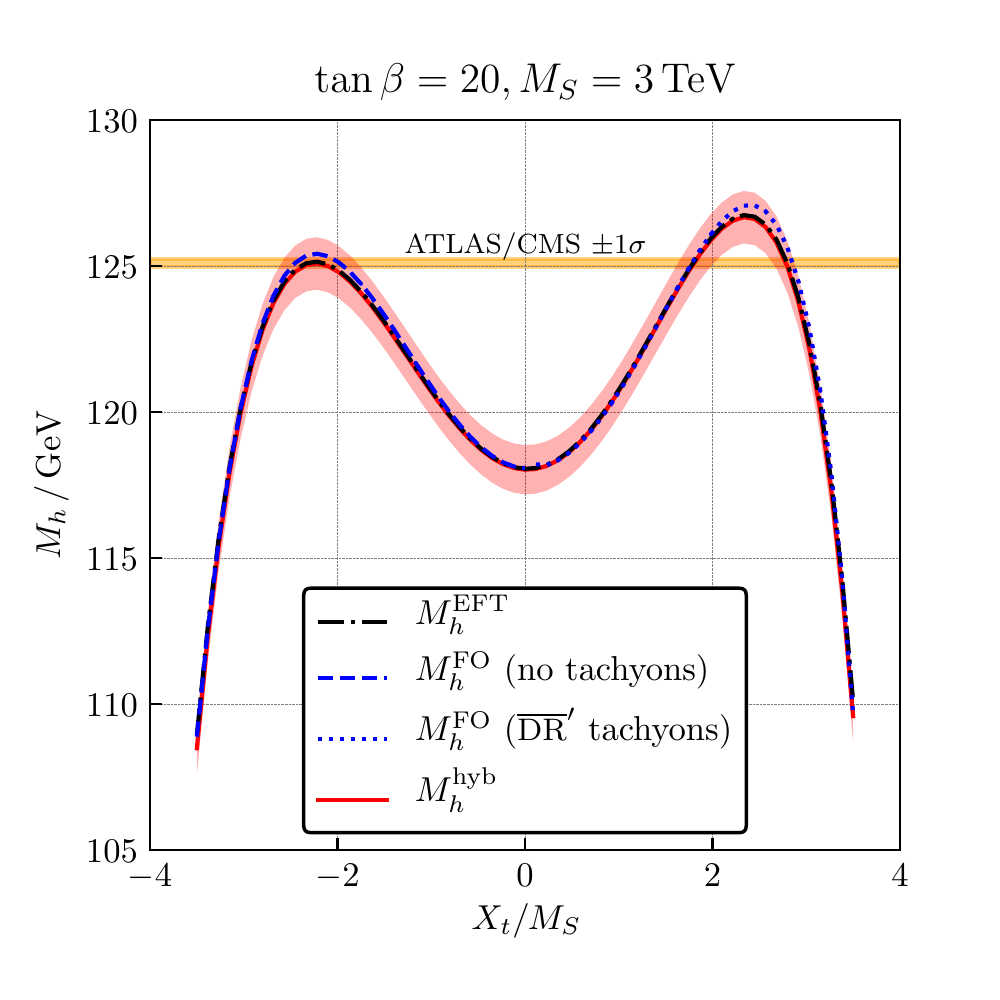}%
    \label{fig:3L_Mh_Xt}%
  }\hfill
  \subfloat[][]{%
    \includegraphics[width=0.49\textwidth]{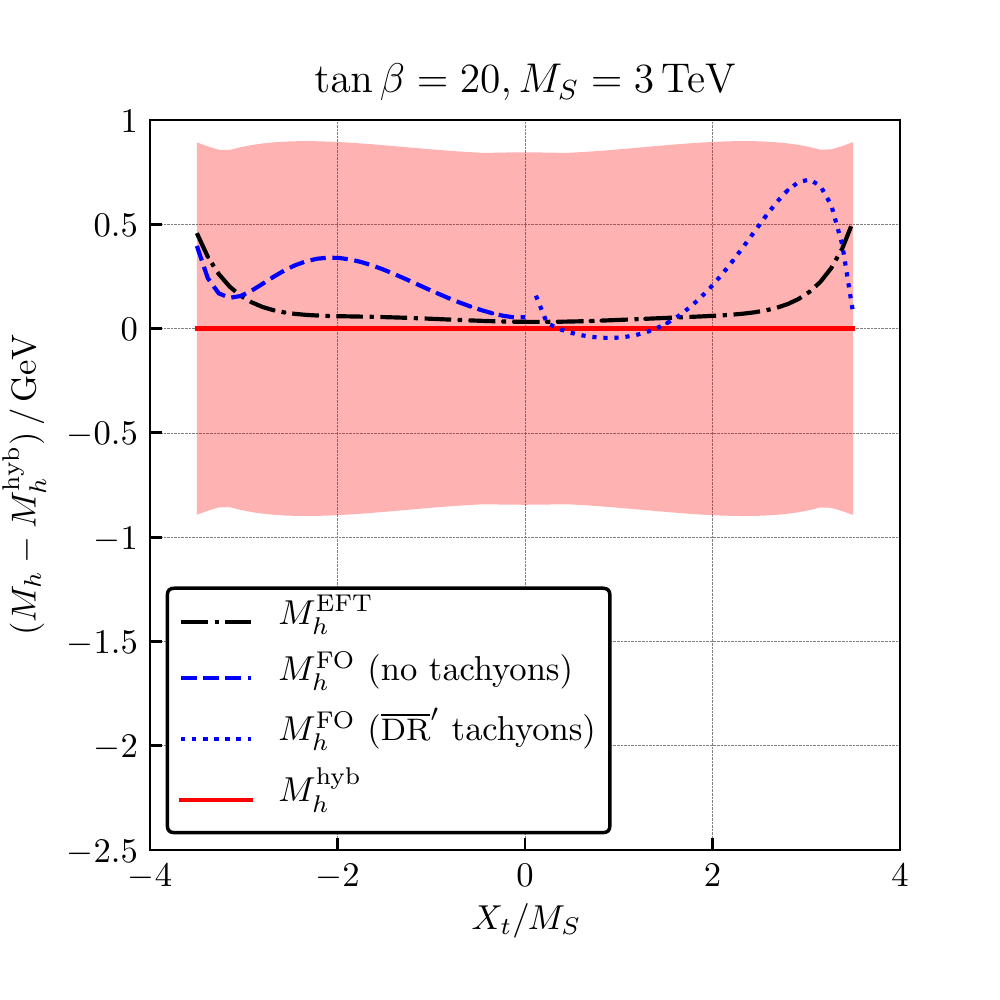}%
    \label{fig:3L_Mh_Xt_diff}%
  }
  \caption{Comparison of the \three-loop \fo,
    \eft, and \mixed\ results as functions of $X_t/\ms$.}
  \label{fig:scan_Xt_mixed_3L}
\end{figure}

In \fig{fig:scan_Xt_mixed_3L} a comparison of the \mixed\ results with
the \three-loop \fo\ and \eft\ ones is shown as a function of $x_t$ for
the degenerate scenario with $\tan\beta = 20$ and $\ms = 3\TeV$, where
the \mssm\ value of the Higgs pole mass can be in agreement with the
experimentally measured value.  As our derivation of the $\Deltasupp$
terms from above suggests, we find agreement of the \mixed\ result with
the \eft\ within $0.5\GeV$ for such a large \susy\ scale.  The largest
deviations of $0.5\GeV$ occur in the region $|x_t| > 3$, while in the
region $|x_t| < 3$ the deviation is smaller than $0.1\GeV$. However,
this region suffers from a problematic feature of the fixed-order
calculation, which is the occurrence of tachyonic \drbarprime\ masses of
the heavy \CP-even, the \CP-odd, and the charged Higgs bosons at the
electroweak scale for $x_t > 0$; this will be discussed in more detail
in \sct{sec:tachyons}.

\subsubsection{Convergence for low \susy\ scales}
\label{sec:non-convergence}

As described in \citeres{Athron:2016fuq,Athron:2017fvs}, the (hybrid)
\FlexibleEFTHiggs\ calculation implemented in \FlexibleSUSY\ since
version 2.0.0 includes all \one-loop contributions and resums all
large logarithmic corrections at the next-to-leading logarithmic level
(\nll).  When compared to the \one-loop fixed-order \drbarprime\
result of \FlexibleSUSY, one finds very good agreement in the limit
$\ms\to M_Z$ if $\tan\beta\to 1$ and $x_t = 0$.  However, for larger
values of $\tan\beta$ or $x_t$ the \FlexibleEFTHiggs\ calculation does
not converge well towards the fixed-order calculation for $\ms\to M_Z$
due to larger incomplete higher-order terms picked up by both
calculations, as can be seen in \fig{fig:scan_MS_mixed_3L}.  In the
following we give examples of sources of such incomplete higher-order
terms.  We start from a scenario with small $\tan\beta$ and $x_t = 0$,
where the incomplete higher-order terms are small.  We then step-wise
increase $\tan\beta$ and $\ms$ and discuss the occurring deviations
between the two calculations.  Note, that the incomplete higher-order
terms are also sensitive to the value of $x_t$.  However, their $x_t$
dependence at the electroweak scale cannot be properly studied for
large values of $x_t$ with the \drbarprime\ calculation implemented in
\FlexibleSUSY\ due to the occurrence of a tachyonic \drbarprime\ stop
mass at $\ms \sim M_Z$.

The first row in \tab{tab:step_by_step} shows the scenario with
$\tan\beta = 3$, $\ms = M_Z$, and $x_t = 0$, where both results agree
within $5\MeV$ ($0.01\%$).
\begin{table}[tbh]
  \centering
  \caption{Comparison of the \one-loop \FlexibleEFTHiggs\ and $n$-loop
    fixed-order \drbarprime\ Higgs pole mass with \FlexibleSUSY.}
  \begin{tabular}{rrrrrrr}
    \toprule
    $n$ & $\tan\beta$ & $\ms$ & $x_t$ & $\MhFEFT$ & $\MhFOFS$ & $(\MhFEFT - \MhFOFS)$ \\
    \midrule
    $1$ & $3$  & $M_Z$     & $0$ & $57.584\GeV$ & $57.590\GeV$ & $-0.005\GeV$ \\
    $1$ & $20$ & $M_Z$     & $0$ & $88.725\GeV$ & $88.636\GeV$ & $+0.089\GeV$ \\
    $1$ & $20$ & $M_t$     & $0$ & $95.612\GeV$ & $95.999\GeV$ & $-0.387\GeV$ \\
    $1$ & $20$ & $200\GeV$ & $0$ & $96.733\GeV$ & $97.378\GeV$ & $-0.645\GeV$ \\
    $1$ & $20$ & $500\GeV$ & $0$ & $105.489\GeV$ & $107.059\GeV$ & $-1.570\GeV$ \\
    $2$ & $20$ & $500\GeV$ & $0$ & $105.489\GeV$ & $105.411\GeV$ & $-0.078\GeV$ \\
    \bottomrule
  \end{tabular}
  \label{tab:step_by_step}
\end{table}
When increasing $\tan\beta$, the \two-loop differences between the two
Higgs mass values become more sizable, increasing to $0.089\GeV$
($0.1\%$) for $\tan\beta=20$, see the second row of
\tab{tab:step_by_step}.  There are multiple sources of such
$\tan\beta$-dependent higher-order terms in both calculations: In the
fixed-order calculation, for example, an iteration over the squared
momentum $p^2$ is used to find the solution of the equation
\begin{align}
  0 = \det\left\{ p^2\delta_{ij} - (m_h^{\mssm})^2_{ij} + \Re\left[\Sigma_h(p^2)_{ij} - \frac{(t_h)_i}{v_i}\delta_{ij}\right]  \right\},
\end{align}
where $\Sigma_h(p^2)$ is the momentum-dependent \CP-even Higgs
self-energy matrix and $t_h$ the tadpole vector (see
\citere{Pierce:1996zz}, for example).  This iteration leads
to higher-order \susy\ contributions of $\order{y_t^ny_b^m v^2/\ms^2}$
($n+m \ge 6$) which increase with $\tan\beta$, for example due to the
increasing bottom Yukawa coupling $y_b$.  In the
\FlexibleEFTHiggs\ approach such terms are absent because $p^2$-terms
are taken into account only at the \one-loop level, and thus no momentum
iteration needs to be performed.  However, in the
\FlexibleEFTHiggs\ calculation other $\tan\beta$-dependent higher-order
terms are generated, for example by inserting the \one-loop threshold
corrections for the \mssm\ \drbarprime\ electroweak gauge couplings
$g_1$ and $g_2$ into the tree-level term $(m_h^{\mssm})^2$ on the
r.h.s.\ of \eqn{eq:FEFT_matching} in order to express the quartic Higgs
coupling of the \sm\ in terms of \sm\ \msbar\ gauge couplings:
\begin{subequations}
    \begin{align}
  (M_h^\mssm)^2 &= (m_h^{\mssm})^2 + \Delta^{1\ell} (m_h^{\mssm})^2, \\
  (m_h^{\mssm})^2 &= \frac{1}{4}\left(\frac{3}{5} g_1^2 + g_2^2\right) c_{2\beta}^2 \, v^2
  \left[ 1 + \left(\frac{3}{5} g_1^2 + g_2^2\right)(c_{2\beta}^2-1) \frac{v^2}{4 m_A^2}
  \right] + \mathcal{O}\left(\frac{v^4}{m_A^4}\right).
\end{align}
\end{subequations}
Since the tree-level \mssm\ \drbarprime\ Higgs mass $(m_h^{\mssm})^2$
initially depends on $g_1^2$, $g_2^2$, and $c_{2\beta} \equiv \cos
2\beta$, the insertion of the threshold corrections generates \two-loop
terms, which are of electroweak order $\order{g_1^n g_2^m
  c_{2\beta}^{2k} v^2/m_A^2}$ and depend on $\tan\beta$.  Note that
these are just two of several possible sources for incomplete
higher-order $\tan\beta$-dependent terms by which the two formally
\one-loop approximations differ.

When the \susy\ scale is increased to $\ms = \Mtop$ (third row in
\tab{tab:step_by_step}), renormalization group running effects come into
play, because the scale at which the running couplings are extracted ($Q
= M_Z$) is no longer identical to the scale where the Higgs pole mass is
calculated at ($Q = \ms = \Mtop$).  While in \FlexibleEFTHiggs\ the
\sm\ \rge{}s are used to evolve the running couplings from $M_Z \to
\Mtop$, the fixed-order calculation uses \mssm\ \rge{}s.  This raises
the difference between the two results to $-0.387\GeV$ ($-0.4\%$) in our
example.
For larger \susy\ scales, this difference increases further, as shown in
4$^\text{th}$ and 5$^\text{th}$ rows of \tab{tab:step_by_step} for
$\ms=200$\,GeV and $\ms=500$\,GeV, respectively. For these scales,
logarithmic corrections of the form $\log(\ms/\Mtop)$ occur, which get
resummed in the \FlexibleEFTHiggs\ calculation, but not in the
fixed-order one.  In the latter, the inclusion of \two-loop corrections
must account for this difference.  In fact, when \two-loop corrections
are included in the fixed-order calculation, see the bottom row of
\tab{tab:step_by_step}, the difference is reduced again to $-0.078\GeV$
($-0.07\%$).

This analysis shows that one cannot expect perfect agreement between the
\FlexibleEFTHiggs\ and the fixed-order results at low \susy\ scales $\ms
\lesssim 200\GeV$, even though both calculations are formally consistent
at their respective accuracy level.  Since the \FlexibleEFTHiggs\ result
is part of our \mixed\ scheme
\eqref{eq:Mh_mixed}--\eqref{eq:Delta_supp_2L}, the described deviation
translates into a non-convergence of $\Mhmixed$ towards the \three-loop
fixed-order result at low \susy\ scales in \fig{fig:scan_MS_mixed_3L}.

\subsubsection{Uncertainty estimate}
\label{sec:uncertainty}

We estimate the uncertainty of the \mixed\ result by taking the minimum
uncertainty of the \fo\ and \eft\ results for each parameter point,
\begin{align}
  \Delta\Mhmixed = \min\left\{ \Delta\MhFO,  \Delta\MhEFT \right\}.
  \label{eq:uncertainty}
\end{align}
The uncertainty of the \three-loop fixed-order calculation,
$\Delta\MhFO$, is estimated by (a) varying the renormalization scale $\mus$ at
which the Higgs pole mass is calculated and (b) by in-/excluding the
\two-loop threshold correction for the strong gauge coupling in the \mssm\
\cite{Harlander:2005wm,Bauer:2008bj,Bednyakov:2010ni}:
\begin{align}
  \Delta\MhFO &= \Delta^{(\mus)} \MhFO + \Delta^{(g_3)} \MhFO,
\end{align}
with
\begin{subequations}
  \begin{align}
  \label{eq:DMhQSa}
  \Delta^{(\mus)} \MhFO &= \max_{\mus\in[\Mtop,\ms]} \left| \MhFO(\mus) - \MhFO(\ms) \right|, \\
  \label{eq:DMhQSb}
  \Delta^{(g_3)} \MhFO &= \left| \MhFO(g_3^{1\ell}) - \MhFO(g_3^{2\ell}) \right|.
\end{align}
\end{subequations}
Even though this uncertainty estimate implicitly assumes a common
\susy\ mass $\ms$, in accordance with our numerical examples in this
paper, its application is not restricted to the exactly degenerate mass
case, of course.  A general \susy\ spectrum may require more
sophisticated estimates of the \fo\ and the \eft\ uncertainties though,
but \eqn{eq:uncertainty} should remain applicable.

We emphasize that the scale variation of $\mus$ in \eqn{eq:DMhQSa} leads
to an enhanced sensitivity of $\Delta^{(\mus)} \MhFO$ to terms of the
order $\order{\log^4(\ms/\Mtop)}$, compared to the corresponding
uncertainty estimates of \citeres{Allanach:2018fif,Athron:2016fuq}, for
example. For \susy\ scales below $0.7$--$0.8\TeV$, the resulting
fixed-order uncertainty is the smaller of the two on the r.h.s.\ in
\eqn{eq:uncertainty}.  Due to the occurrence of large logarithmic loop
corrections, $\Delta\MhFO$ becomes larger when $\ms$ is increased and
reaches about $\Delta\MhFO \approx 2\GeV$ for $\ms \approx 0.7\TeV$ and
$x_t=-\sqrt{6}$.

The uncertainty of the \three-loop \eft\ calculation, $\Delta\MhEFT$, is
estimated by (a) varying the renormalization scale $\Qpole$ at which the
Higgs pole mass is calculated, (b) varying the renormalization scale
$\Qmatch$ at which the \mssm\ is matched to the \sm, (c) ex-/including
the \four-loop \qcd\ threshold correction for the \sm\ top Yukawa
coupling \cite{Martin:2016xsp}, and (d) estimating the effect of
$\order{v^2/\ms^2}$ terms from the quartic Higgs coupling along the
lines of
\citeres{Bagnaschi:2014rsa,Vega:2015fna,Allanach:2018fif}:\footnote{I.e.\
  $\MhEFT(v^2/\ms^2)$ of \eqn{eq:efterrd} is obtained by scaling the
  individual terms in the \one-loop threshold correction
  $\Delta\lambda^{1\ell}$ for the quartic coupling by factors of the
  order $(1+v^2/\ms^2)$.}
\begin{align}
  \Delta\MhEFT &= \Delta^{(\Qpole)} \MhEFT + \Delta^{(\Qmatch)} \MhEFT + \Delta^{(y_t^{\sm})} \MhEFT + \Delta^{(v^2/\ms^2)} \MhEFT, \label{eq:DMh_EFT}
\end{align}
with
\begin{subequations}
  \begin{align}
  \Delta^{(\Qpole)} \MhEFT &= \max_{Q\in[\Mtop/2,2\Mtop]} \left| \MhEFT(Q) - \MhEFT(\Mtop) \right|, \\
  \Delta^{(\Qmatch)} \MhEFT &= 0.5\GeV, \label{eq:qmatcherr}\\
  \Delta^{(y_t^{\sm})} \MhEFT &= \left| \MhEFT(y_t^{\sm,3\ell}) - \MhEFT(y_t^{\sm,4\ell}) \right|, \\
  \Delta^{(v^2/\ms^2)} \MhEFT &= \left| \MhEFT - \MhEFT(v^2/\ms^2)
  \right|.
  \label{eq:efterrd}
\end{align}
\end{subequations}
$\Delta^{(\Qpole)} \MhEFT$ is approximately independent of the
\susy\ scale and amounts to about $0.2\GeV$.
The matching scale uncertainty $\Delta^{(\Qmatch)} \MhEFT$ has been
estimated in
\citeres{Bagnaschi:2014rsa,Vega:2015fna,Allanach:2018fif}. It was found
that for scenarios as those considered here, the uncertainty does not
exceed 0.5\,GeV for $\ms \gtrsim 1\TeV$. A generalization of this
uncertainty estimate to our result would require the extension of the
underlying procedure to \nkll{3}, which would involve the logarithmic
terms at \nklo{4} and their implementation into \HSSUSY. As long as this
is not available, we content ourselves to conservatively associate the
maximal value of 0.5\,GeV (see above) with the matching scale
uncertainty, independent of $\ms$.
For the uncertainty $\Delta^{(y_t^{\sm})} \MhEFT$ induced by the
\sm\ top Yukawa coupling we follow the prescription of
\citeres{Vega:2015fna,Allanach:2018fif}, but apply it at the next order
in perturbation theory as required by our results. It amounts to
approximately $0.1\GeV$ and increases slightly with the \susy\ scale.
For \susy\ scales above $1$--$2\TeV$, the total uncertainty of the
\eft\ calculation $\Delta\MhEFT$ is dominated by these three
contributions and amounts to slightly less than $1\GeV$, while
$\Delta^{(v^2/\ms^2)}\MhEFT$ is negligible.  This is in agreement with
the results from \sct{sec:supp}, where it was found that the
$\order{v^2/\ms^2}$ terms are below $0.25\GeV$ for $\ms\gtrsim 1\TeV$.
Finally, we find that the uncertainty $|\delta_{x_t} +
\delta_{\text{exp}}|$ of the \three-loop calculation of $\lambda$ given
in \citere{Harlander:2018yhj} is below $2\MeV$ for the degenerate-mass
scenarios considered here with $\ms \gtrsim 1\TeV$, and is thus
negligible.

Our combined (hybrid) uncertainty \eqref{eq:uncertainty} is shown as red
band in \figs{fig:scan_MS_mixed_3L}--\ref{fig:scan_Xt_mixed_3L}.  For
degenerate mass scenarios with $\tan\beta = 20$, the uncertainty for
large $\ms$ is fairly constant and slightly below $1$\,GeV. Part of this
behavior is implied by our constant choice for $\Delta^{(\Qmatch)}\MhEFT$, of
course, but this accounts only for about 60\% of the full band in this
region. The uncertainty increases towards lower $\ms$, according to the
decreasing accuracy of the \eft\ approach, until \eqn{eq:uncertainty}
switches to the \fo\ uncertainty to determine $\Delta\Mhmixed$. This
happens around $\ms\sim 0.7$--$0.8\GeV$, where the hybrid uncertainty
reaches up to $2$--$2.5\GeV$ for large values of $x_t$. Note that the
numerical values of the \fo\ and the \eft\ result are compatible with
each other in this region, which underlines the validity of our
approach.  Further decreasing $\ms$ leads to a significant improvement
of the hybrid uncertainty, reflecting the increasing reliability of
the \fo\ approach in low-scale \susy\ scenarios.

Very rarely it happens that the central value of the approach (\eft\ or
\fo) that determines the hybrid uncertainty through \eqn{eq:uncertainty}
is not itself contained in the resulting uncertainty band. In this case,
we widen the band correspondingly.

Quite generally, we find that the \susy\ scale $\msequal$, where both
the \fo\ and the \eft\ calculation have the same uncertainty, is
between $\ms \sim 0.7$--$0.8\TeV$.  This region is slightly lower than
our estimate from \sct{sec:supp}.  Due to our different uncertainty
estimate of the fixed-order calculation, this region is also below the
region of $\msequal = 1.0$--$1.3\TeV$ estimated in
\citere{Allanach:2018fif}.  For larger values of $\ms$ the \eft\
calculation deviates not more than $0.5\GeV$ from the hybrid
calculation.  The behavior of the curves and the associated
uncertainty also justifies our estimate with hindsight. For example,
the \fo\ result visibly impacts the \mixed\ result only for rather low
values of $\ms\lesssim 1$\,TeV, but there the interval $[\Mtop,\ms]$
for the renormalization scale variation is sufficiently close to the
weak scale to provide a reasonable estimate of the theory uncertainty.

\subsection{Tachyonic Higgs bosons at the electroweak scale}
\label{sec:tachyons}

\begin{figure}[tbh]
  \centering
  \subfloat[][]{%
    \includegraphics[width=0.49\textwidth]{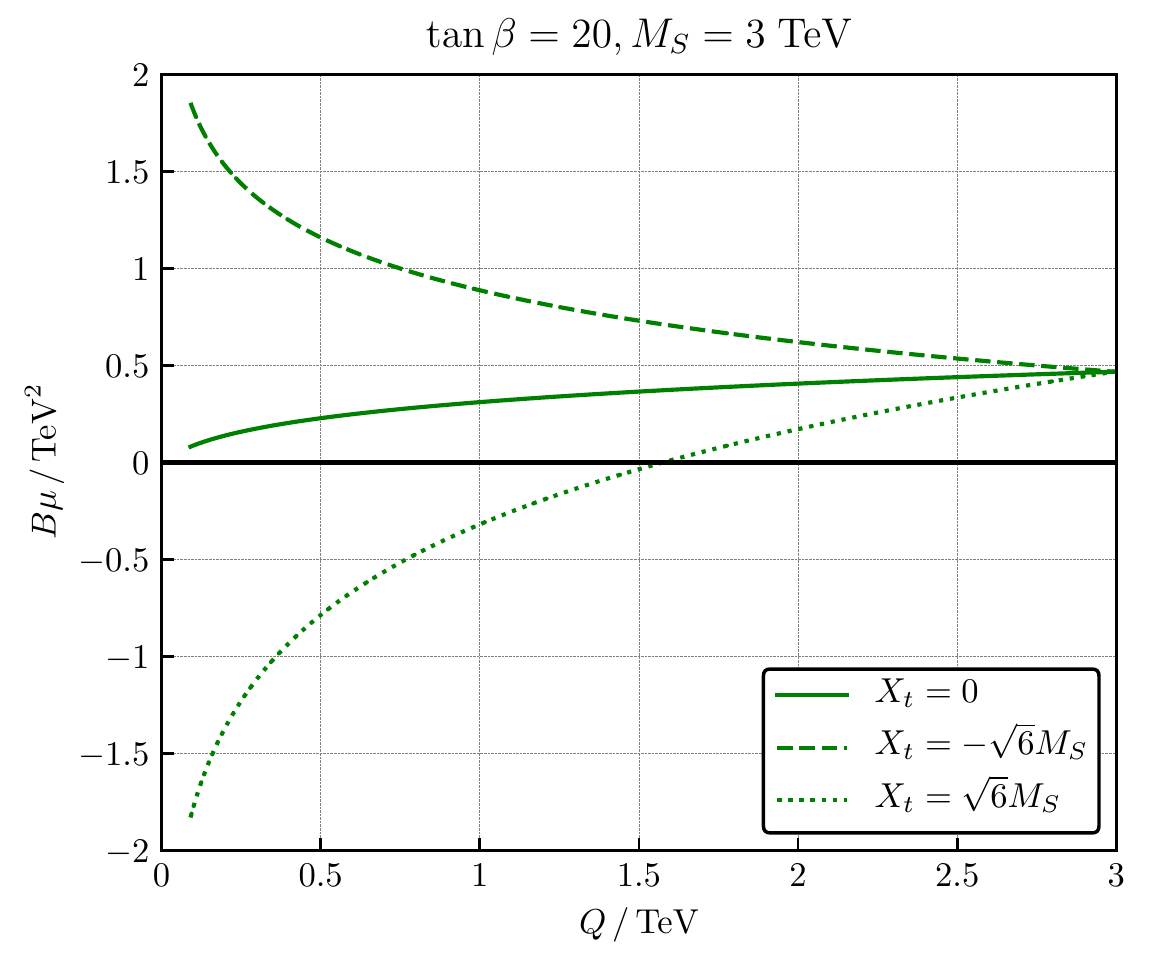}%
    \label{fig:3L_BMu_Q}%
  }\hfill
  \subfloat[][]{%
    \includegraphics[width=0.49\textwidth]{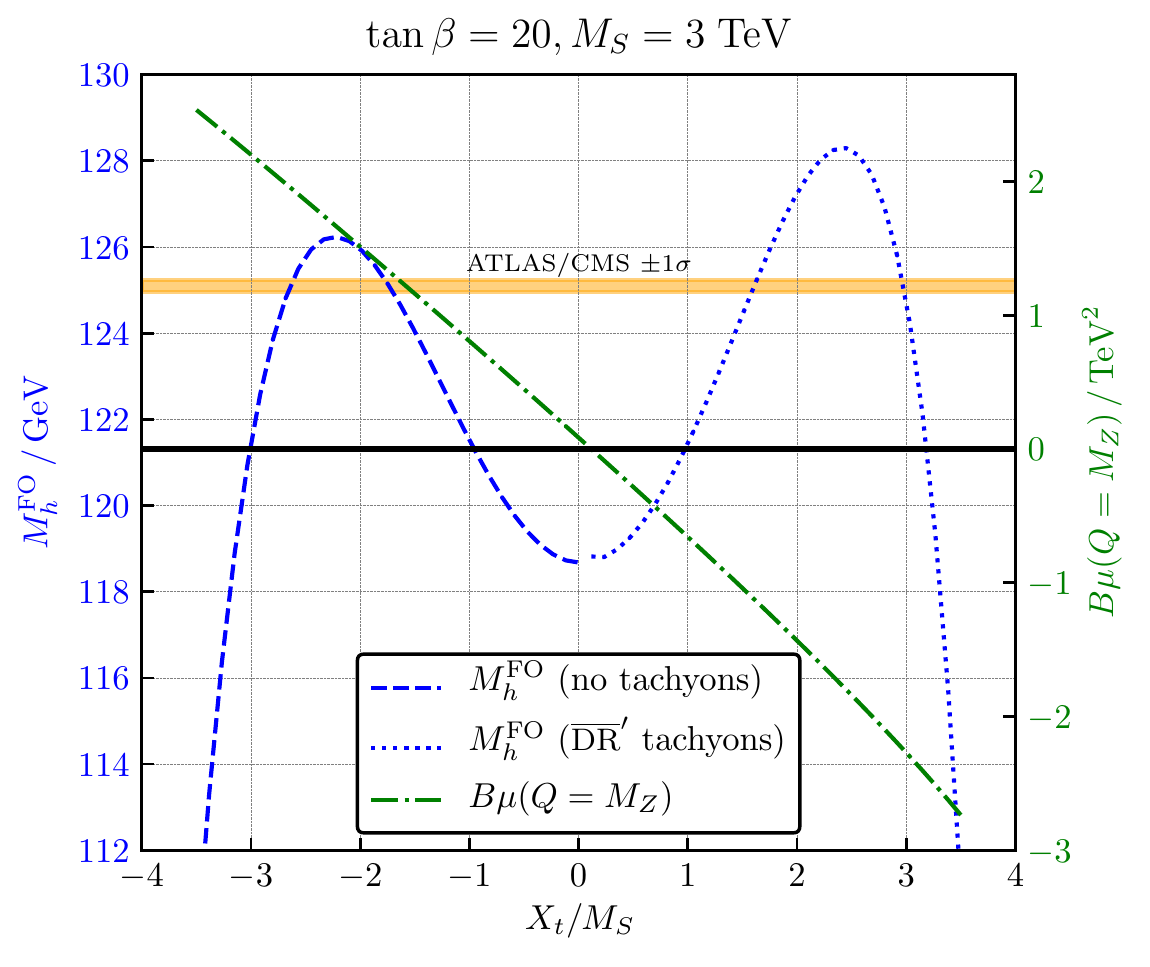}%
    \label{fig:3L_Mh_Xt_tachyons}%
  }
  \caption{Left panel: Renormalization group running of $B\mu(Q)$ for
    different values of $X_t$.  Right panel: Three-loop fixed-order
    Higgs pole mass (blue lines) and $B\mu(Q=M_Z)$ as a
    function of $X_t/\ms$ (green dash-dotted line).}
  \label{fig:scan_Xt_tachyons}
\end{figure}
As described in the previous section, in the fixed-order calculation the
\drbarprime\ masses of the heavy \CP-even, the \CP-odd, and the charged
Higgs bosons are tachyonic at the scale $Q=M_Z$ for $x_t \gtrsim 0$.
The reason for this is the $B\mu$ parameter, which is negative at that
scale due to the renormalization group running, see \fig{fig:3L_BMu_Q}.
In our scenario, the value of $B\mu$ is fixed at the \susy\ scale by the
\drbarprime\ \CP-odd Higgs mass $m_A(\ms)$ as
\begin{align}
  B\mu(\ms) = \frac{1}{2} \sin[2\beta(\ms)] \, m_A^2(\ms) \approx 0.05 \ms^2,
\end{align}
where we have set $\tan\beta(\ms) = 20$ and $m_A^2(\ms) = \ms^2$ in the last
step.  For such a large value of $\tan\beta$, the \one-loop
$\beta$-function of the $B\mu$ parameter is approximately given by
\begin{align}
   \beta_{B\mu} &\approx 3 \kappa y_t^2 \left( B\mu + 2 \mu A_t \right)
  \approx 3 \kappa y_t^2 \left( 0.05 + 2 x_t \right) \ms^2.
\end{align}
For $x_t < -0.025$ the $\beta$-function is negative, which means that
$B\mu$ increases during the renormalization group running from $\ms$
down to $M_Z$, see the green dashed line in \fig{fig:3L_BMu_Q}.
However, if $x_t > -0.025$ the $\beta$-function is positive so that
$B\mu$ decreases when running down and changes sign at some low scale
$Q_\text{tach}$ (green dotted line).  The value of the scale
$Q_\text{tach}$ can be larger than $M_Z$ if $x_t$ and/or $\ms$ are large
enough, for example for $x_t > 0$ and $\ms \gtrsim 3\TeV$.  When this
happens, the \drbarprime\ masses of the heavy \CP-even, the \CP-odd, and
the charged Higgs bosons are tachyonic at $Q=M_Z$, because
\begin{align}
  m_H^2(M_Z) \approx m_{H^{\pm}}^2(M_Z) \approx m_A^2(M_Z) = \frac{2 B\mu(M_Z)}{\sin[2\beta(M_Z)]} < 0.
\end{align}
In \fig{fig:3L_Mh_Xt_tachyons} the value of $B\mu(M_Z)$ is shown as a
function of $x_t$ as green dash-dotted line for the scenario with
$\tan\beta = 20$ and $\ms = 3\TeV$. In accordance with the estimate
above, $B\mu(M_Z)$ is in fact negative for positive values of $x_t$, and
the \fo\ Higgs mass calculation (blue dashed/dotted lines) involves
tachyonic \drbarprime\ masses at the electroweak scale. At the very
least, this implies that the loop corrections to the heavy Higgs boson
masses are very large.  In some spectrum generators, the occurrence of
heavy Higgs tachyons is bypassed by using the \textit{pole} masses of
the heavy Higgs boson masses in the loop calculations at the low scale,
instead of the \drbarprime\ masses.  In \FlexibleSUSY, on the other
hand, an error is flagged by default if \drbarprime\ tachyons appear at
any scale. Optionally, \FlexibleSUSY\ uses the absolute values of the
tachyonic masses in the loop integrals.\footnote{This is achieved by
  setting the flag \code{FlexibleSUSY[12] = 1} in the \slha\ input or
  \code{forceOutput -> 1} in the \Mathematica\ interface.}  This option
was used in \figs{fig:scan_Xt_mixed_3L} and \ref{fig:3L_Mh_Xt_tachyons}
for $x_t>0$, which explains the kink at $x_t = 0$ in the fixed-order
curve in \fig{fig:3L_Mh_Xt_diff}, as replacing negative by positive
squared masses is not a smooth transition.

In general, the occurrence of these tachyonic states due to higher order
effects appears to make the approach \cite{Pierce:1996zz} of matching
\sm\ and \mssm\ parameters at the scale $M_Z$ questionable.  For
\susy\ scales above the TeV scale it might thus be advisable to perform
the matching at a larger scale to avoid tachyonic states.  To our
knowledge, this program has not been pursued in all generality up to now
(see \citere{Kunz:2014gya}, however).  For very large \susy\ scales, the
\fo\ approach is bound to fail anyway due to the large logarithms as
discussed in the introduction.

\section{Conclusions}
\label{sec:conclusions}

We presented a hybrid calculation of the light \CP-even Higgs boson pole
mass in the real \mssm\ by combining \fo\ and \eft\ results.  Our
procedure is based on the \drbarprime\ scheme. Beyond the relevant
two-loop \fo\ corrections and the corresponding resummation of large
logarithms through \nnll, our result includes the \three-loop
\fo\ corrections and the resummation through \nkll{3} w.r.t.\ the strong
coupling.

The estimated uncertainty of our hybrid result is below $1\GeV$ in
most of the relevant parameter space.  An exception is the
transition region $\ms = 0.7$--$2\TeV$, where both the fixed-order
and the \eft\ calculation are less precise and the uncertainty can be
up to $\sim 2\GeV$.

By comparing the hybrid calculation with the pure \eft\ calculation, we
can estimate the size of the terms of $\order{v^2/\ms^2}$ which are
typically neglected in a pure \eft\ approach.  For degenerate
\susy\ mass parameters we find that these terms are smaller than $0.25
\GeV$ as long as $\ms\gtrsim 1\TeV$, which is the region where the
degenerate scenarios can be compatible with the experimental value for
the Higgs mass \cite{Bagnaschi:2014rsa}.  Combining this with the fact
that for $\ms\gtrsim 0.7$--$0.8$$\TeV$ the pure \eft\ calculation has a
smaller uncertainty than the \fo\ calculation (see also
\citere{Allanach:2018fif}), we conclude that a pure \eft\ calculation
provides an excellent approximation in the \mssm\ for the degenerate
\susy\ mass parameter scenarios.

\section*{Acknowledgments}

We would like to thank Lars-Thorben Moos for collaboration at early
states of this project, and Thomas Kwasnitza and Dominik St\"ockinger
for helpful discussions about the \FlexibleEFTHiggs\ approach.  This
research was supported by the DFG Collaborative Research Center
``Particle Physics Phenomenology after the Higgs Discovery'' (TRR
257), and the Research Unit ``New Physics at the LHC'' (FOR 2239).

\def\app#1#2#3{{\it Act.~Phys.~Pol.~}\jref{\bf B #1}{#2}{#3}}
\def\apa#1#2#3{{\it Act.~Phys.~Austr.~}\jref{\bf#1}{#2}{#3}}
\def\annphys#1#2#3{{\it Ann.~Phys.~}\jref{\bf #1}{#2}{#3}}
\def\cmp#1#2#3{{\it Comm.~Math.~Phys.~}\jref{\bf #1}{#2}{#3}}
\def\cpc#1#2#3{{\it Comp.~Phys.~Commun.~}\jref{\bf #1}{#2}{#3}}
\def\epjc#1#2#3{{\it Eur.\ Phys.\ J.\ }\jref{\bf C #1}{#2}{#3}}
\def\fortp#1#2#3{{\it Fortschr.~Phys.~}\jref{\bf#1}{#2}{#3}}
\def\ijmpc#1#2#3{{\it Int.~J.~Mod.~Phys.~}\jref{\bf C #1}{#2}{#3}}
\def\ijmpa#1#2#3{{\it Int.~J.~Mod.~Phys.~}\jref{\bf A #1}{#2}{#3}}
\def\jcp#1#2#3{{\it J.~Comp.~Phys.~}\jref{\bf #1}{#2}{#3}}
\def\jetp#1#2#3{{\it JETP~Lett.~}\jref{\bf #1}{#2}{#3}}
\def\jphysg#1#2#3{{\small\it J.~Phys.~G~}\jref{\bf #1}{#2}{#3}}
\def\jhep#1#2#3{{\small\it JHEP~}\jref{\bf #1}{#2}{#3}}
\def\mpl#1#2#3{{\it Mod.~Phys.~Lett.~}\jref{\bf A #1}{#2}{#3}}
\def\nima#1#2#3{{\it Nucl.~Inst.~Meth.~}\jref{\bf A #1}{#2}{#3}}
\def\npb#1#2#3{{\it Nucl.~Phys.~}\jref{\bf B #1}{#2}{#3}}
\def\nca#1#2#3{{\it Nuovo~Cim.~}\jref{\bf #1A}{#2}{#3}}
\def\plb#1#2#3{{\it Phys.~Lett.~}\jref{\bf B #1}{#2}{#3}}
\def\prc#1#2#3{{\it Phys.~Reports }\jref{\bf #1}{#2}{#3}}
\def\prd#1#2#3{{\it Phys.~Rev.~}\jref{\bf D #1}{#2}{#3}}
\def\pR#1#2#3{{\it Phys.~Rev.~}\jref{\bf #1}{#2}{#3}}
\def\prl#1#2#3{{\it Phys.~Rev.~Lett.~}\jref{\bf #1}{#2}{#3}}
\def\pr#1#2#3{{\it Phys.~Reports }\jref{\bf #1}{#2}{#3}}
\def\ptp#1#2#3{{\it Prog.~Theor.~Phys.~}\jref{\bf #1}{#2}{#3}}
\def\ppnp#1#2#3{{\it Prog.~Part.~Nucl.~Phys.~}\jref{\bf #1}{#2}{#3}}
\def\rmp#1#2#3{{\it Rev.~Mod.~Phys.~}\jref{\bf #1}{#2}{#3}}
\def\sovnp#1#2#3{{\it Sov.~J.~Nucl.~Phys.~}\jref{\bf #1}{#2}{#3}}
\def\sovus#1#2#3{{\it Sov.~Phys.~Usp.~}\jref{\bf #1}{#2}{#3}}
\def\tmf#1#2#3{{\it Teor.~Mat.~Fiz.~}\jref{\bf #1}{#2}{#3}}
\def\tmp#1#2#3{{\it Theor.~Math.~Phys.~}\jref{\bf #1}{#2}{#3}}
\def\yadfiz#1#2#3{{\it Yad.~Fiz.~}\jref{\bf #1}{#2}{#3}}
\def\zpc#1#2#3{{\it Z.~Phys.~}\jref{\bf C #1}{#2}{#3}}
\def\ibid#1#2#3{{ibid.~}\jref{\bf #1}{#2}{#3}}
\def\otherjournal#1#2#3#4{{\it #1}\jref{\bf #2}{#3}{#4}}
\newcommand{\jref}[3]{{\bf #1}, #3 (#2)}
\newcommand{\hepph}[1]{\href{https://arXiv.org/abs/hep-ph/#1}{\texttt{hep-ph/#1}}}
\newcommand{\mathph}[1]{{\tt [math-ph/#1]}}
\newcommand{\arxiv}[2]{\href{https://arXiv.org/abs/#1}{\texttt{arXiv:#1\,[#2]}}}
\newcommand{\bibentry}[4]{#1, {\it #2}, #3\ifthenelse{\equal{#4}{}}{}{, }#4.}

\end{document}